\documentstyle[amssymb,amstex,marvosym,epsfig,12pt]{article}

\def\beq{\begin{equation}}
\def\eeq{\end{equation}}
\def\bea{\begin{eqnarray}}
\def\eea{\end{eqnarray}}
\def\bq{\begin{quote}}
\def\eq{\end{quote}}

\def\bq{\begin{quote}}
\def\eq{\end{quote}}

\begin{document}
\titlepage                                                          
\begin{flushright}                                                     
OUTP-02-37P\\                                                       
\end{flushright} 
\begin{center}                                                        
{\Large \bf Symmetries and fermion masses}\\ 
\vspace*{0.5cm}                                                          
G.\ G.\ Ross{\footnote{g.ross1$\MVAt$physics.ox.ac.uk}} and L.\ Velasco-Sevilla{\footnote{velasco$\MVAt$thphys.ox.ac.uk}} \\              Department of Physics, Theoretical Physics, University of Oxford,\\
1 Keble Road, Oxford OX1 3NP,U.K.\\         
\end{center}   

\begin{abstract}
We discuss whether quark, charged lepton and neutrino masses and mixing
angles may be related by an extended flavour and family symmetry group. We
show that current measurements of all fermion masses and mixing angles are
consistent with a combination of an underlying $SU(3)$ family symmetry
together with a GUT symmetry such as $SO(10)$. In this the near bi-maximal
mixing observed in the neutrino sector is directly related to the small
mixing observed in the quark sector, the difference between quark and lepton
mixing angles being due to the see-saw mechanism. Using this connection we
make a detailed prediction for the lepton mixing angles determining neutrino
oscillation phenomena.
\end{abstract}

\section{Introduction}

While the measurement of quark and lepton masses and mixing angles continues
to improve, their theoretical understanding remains elusive. In the Standard
Model the quark masses and mixing angles and charged lepton masses are
determined by the Yukawa couplings, parameters which must be specified when
defining the Standard Model. The renormalisable couplings of the Standard
Model do not allow neutrino masses although they can be introduced through
the addition of higher dimensional operators, presumably originating in
physics Beyond the Standard Model.

There are a few promising suggestions for structure Beyond the Standard
Model which address the fermion mass problem. For simple symmetry breaking
schemes Grand Unification can relate quark and lepton masses. The most
promising of such relations is the equality $m_{b}=m_{\tau },$ a result
which applies at the GUT scale. Radiative corrections are dominated by QCD
interactions which increase the bottom quark mass at low energy scales,
giving a prediction for the bottom quark mass in good agreement with
experiment. A fairly straightforward generalisation of such GUT relations,
which invokes a new family symmetry, also provides good relations between
the down quarks and charged leptons of the first two generations \cite
{Georgi:1979df}.

However the recent measurement of neutrino masses and mixing angles has shed
doubt on the validity of simple relations between quark and lepton masses
and mixing angles. The most obvious difference is the fact that neutrinos
have masses much smaller than those of the quarks or charged leptons. In the
context of Grand Unification the ``see-saw'' mechanism \cite{GellRamSla79, Yanag79, Mohapatra:1980ia} provides an elegant
mechanism to explain this difference. In this scheme the Standard Model
spectrum is extended to include the right-handed ($SU(2)$ singlet) partners
of the SM neutrinos. As these states can acquire $SU(3)\times SU(2)\times
U(1)$ \ invariant Majorana masses, naturalness implies they should acquire
mass at the scale (the GUT scale?) at which the theory Beyond the Standard
Model breaks to the Standard Model. In addition one expects Dirac masses
between the left- and right- handed neutrino states. These are generated by
electroweak symmetry breaking in a similar way to that of the charged
leptons and quarks and so it is reasonable to assume they will be of
comparable magnitude. Diagnosing the matrix of Dirac and Majorana masses
gives the mass matrix for the three light states, dominantly composed of the
doublet neutrinos, of the form 
\begin{equation}
M_{eff}=M_{D}^{\nu }.M_{\nu }^{-1}.M_{D}^{\nu \text{T}}  \label{seesaw}
\end{equation}
Here $M_{\nu }$ is the $3\times 3$ matrix of Majorana masses for the three
generations of right-handed neutrinos and $M_{D}^{\nu }$ is the $3\times 3$
matrix of Dirac neutrino masses. Given that the natural size of $M_{D}^{\nu
} $ is much less than that of $M_{\nu },$ it is clear this offers an elegant
explanation for the smallness of neutrino masses.

There remains the question why lepton mixing angles should be so different
from quark mixing angles. This is the question we address here in the light
of new information about the quark mass matrix that has recently been
obtained from an analysis of the wealth of b-quark data and the information from neutrino oscillations. Our approach is to
look for the largest symmetry consistent with the fermion masses
concentrating on the possibility that quark and lepton masses are related.
We show that the data in the quark sector data is consistent with an
underlying (spontaneously broken) non-Abelian family symmetry together with
an extension of the Standard Model flavour symmetry to include an $SU(2)_{R}$
flavour symmetry acting on the right-handed components. The charged lepton
masses and neutrino masses and mixings are consistent with this symmetry
provided the flavour symmetry is extended, for example to $SO(10)$, to
generate GUT relations between lepton and quarks. This structure can lead
quite naturally to large lepton mixing angles. Indeed, due to the see-saw
mechanism, lepton mixing angles are large because quark mixing angles are
small \cite{Ross:2000fn, kingross}!

\section{The quark mass matrix \label{quarkmassmats}}

Our starting point is the form of the quark mass matrix. Since experiment is
only able to determine the mass eigenvalues and the CKM mixing matrix, it is
not possible unambiguously to determine the quark mass matrix because the
full form of the left- handed and right-handed rotation matrices needed to
diagonalise the quark masses is not known. However the very reasonable
assumption that the smallness of the mixing angles is due to the smallness
of the mixing angles in the up- and down- sectors separately, allows one to
determine the mass matrix elements on and above the diagonal to good
precision for the down quarks and to lesser precision for the up quarks. In
reference \cite{rrrv}, under this assumption, the quark mass matrices were
determined using the most recent experimental data. This gave the form 
\begin{equation}
\frac{M^{u}}{m_{t}}=\left( 
\begin{array}{ccc}
0 & b^{\prime }\epsilon ^{3} & c^{\prime }\epsilon ^{3} \\ 
b^{\prime }\epsilon ^{3} & \epsilon ^{2} & a^{\prime }\epsilon ^{2} \\ 
? & ? & 1
\end{array}
\right)  \label{mu}
\end{equation}

and 
\begin{equation}
\frac{M^{d}}{m_{b}}=\left( 
\begin{array}{ccc}
0 & b\bar{\epsilon}^{3} & c\bar{\epsilon}^{3} \\ 
b\bar{\epsilon}^{3} & \bar{\epsilon}^{2} & a\bar{\epsilon}^{2} \\ 
? & ? & 1
\end{array}
\right)  \label{md}
\end{equation}
In this the parameters of the up quark mass matrix $\ $are given by $%
\epsilon =0.05,$ $b^{\prime }\simeq e^{i\phi },$ where in the phase
convention used here $\phi \approx 90^{0}$ is the `standard' (i.e.PDG
convention) CP violating phase $\delta $ \cite{rrrv} while $a^{\prime }$ and 
$c^{\prime }$ are very weakly constrained. The parameters of the down quark
mass matrix are much better determined with 
\begin{eqnarray}
\bar{\epsilon} &=&0.15\pm 0.01\quad b=1.5\pm 0.1\quad a=1.31\pm 0.14 
\nonumber \\
{|c|} &=&{0.4\pm 0.05\quad \psi \equiv Arg(c)=-24^{0}\pm 3^{0}}\quad \text{ or }
\nonumber \\
{|c|} &=&1.28{\pm 0.05\quad \psi =60^{0}\pm 5^{0}}\quad \label{fit}
\end{eqnarray}
In both $M^u$ and $M^{d}$ the entries marked with a question mark are
only weakly constrained. In what follows we shall consider the case that the
mass matrices are symmetric or antisymmetric or some combination of the two
so that the elements below the diagonal are also determined. This is in the
spirit of looking for the maximal symmetry consistent with the masses and is
motivated by certain GUTs, notably $SO(10).$ There is one strong piece of
experimental evidence for such a symmetric structure, namely the success of
the Gatto, Sartori, Tonin relation \cite{gst} between the Cabibbo angle and
the quark masses of the first two generations 
\begin{equation}
V_{us}=\sqrt{\frac{m_{d}}{m_{s}}}-\sqrt{\frac{m_{u}}{m_{c}}}e^{i\sigma }
\label{gst}
\end{equation}
where $\sigma $ is the CP violating phase entering the Jarlskog invariant 
\cite{rrrv,Fritzsch:1999ee}. This relation only applies if
the $(1,2)$ and $(2,1)$ matrix elements are equal in magnitude. The solution
of eqs(\ref{mu}) and (\ref{md}) has this structure. Of course there is no
direct indication that the same reflection symmetry applies to the remaining
matrix elements but it is the most natural generalisation of this result. In
any case, as stressed above, we think it of interest to ask whether the most
symmetric form for the quark mass matrices can be simply related to viable
forms for the lepton mass matrices.

\section{Quark family and flavour symmetries\label{family}}

In this Section we will discuss whether the quark mass matrices are
consistent with a larger set of symmetries than those of the Standard Model%
\footnote{%
Here we shall only consider supersymmetric theories.}. These may be flavour
symmetries acting in the same way on each of the three families; for example
simple Grand Unified theories are flavour symmetries. Alternatively they may
also be family symmetries distinguishing between families. In our opinion
the structure of the quark mass matrices strongly suggests an underlying
(spontaneously) broken family symmetry. In such a scheme, in leading order,
only the $(3,3)$ elements are allowed. In terms of the parameter, $\epsilon
, $ \ characterising the symmetry breaking the remaining matrix elements are
filled in at some order $\epsilon ^{n}$ which is also determined by the
symmetry, thus generating the hierarchical structure of fermion masses.

The candidate family symmetry group, namely those symmetries which commutes
with the gauge interactions of the Standard Model, is quite large. For the
quarks it is $U(3)^{3},$ corresponding to a separate $U(3)$ factor for the
left-handed doublets and each of the right-handed $SU(2)$ singlet fields
respectively. To generate quark mass matrices with a left- right- symmetry
we restrict the choice to one with the same transformation for the
left-handed and the charge conjugate right-handed quarks, this symmetry is
reduced to a diagonal $U(3)$ or one of its subgroups\footnote{%
It is possible there is an even larger symmetry with separate gauge group
factors operating on left-and right- handed components and related by a
discrete reflection symmetry. If one requires the left- right- symmetry is
preserved on spontaneous symmetry breaking, the phenomenological
implications of this scheme are the same as the case considered here.}.

Abelian family symmetries have been widely explored. They are capable of
explaining the hierarchical structure of the quark masses and generate quite
acceptable forms for the quark matrices \cite{Ibanez:1994ig,Ross:2000fn}.
The main shortcoming of these schemes is that they do not give a reason for
the smallness of $V_{cb}.$ This follows because if one assigns Abelian
charges to generate the mass hierarchy between $m_{s}$ and $m_{b},$ it leads
to the prediction for $V_{cb}\simeq \sqrt{m_{s}/m_{b}}$ which is too large 
\cite{Ibanez:1994ig,Lola:1999un}. Although this relation is quite sensitive
to corrections of $O(1)$ due to unknown Yukawa couplings, it is an
unsatisfactory feature of an Abelian symmetry. To obtain the correct form
for $V_{cb}$ it is necessary for the ratio of the $(2,2)$ and $(2,3)$ matrix
elements to be close to $1;$ a ratio of $1$ implies $V_{cb}\simeq $ $%
m_{s}/m_{b},$ in good agreement with experiment. In contrast to the Abelian
case a non-Abelian family symmetry can relate the magnitude of the couplings
in different positions in the mass matrix and thus offers the possibility of
explaining this near equality.

\subsection{Non-Abelian family symmetry\label{nonabelian}}

The largest non-Abelian symmetry which treats the left- and charge conjugate
right- handed components in the same way is $SU(3).$ In \cite{kingross} a
specific $SU(3)$ theory has been constructed which relates the $(2,2)$ and $%
(2,3)$ matrix elements. In it the left-handed quark doublets and the up and
down charge conjugate quarks are assigned to $SU(3)$ triplets, $\psi _{i},$ $%
u_{i}^{c}$ and $d_{i}^{c}$ respectively. The critical part of the theory
lies in the pattern of symmetry breaking which leads to the desired mass
matrix. This is achieved through the antitriplet scalar fields $\phi
_{23}^{i},$ $\phi _{3}^{i}$ and triplet fields $\overline{\phi }_{23,i},$ $%
\overline{\phi }_{3,i}$ which acquire vacuum expectation values (VEVs) $\phi
_{3}^{T}=\left( 
\begin{array}{ccc}
0 & 0 & M
\end{array}
\right) $, $\phi _{23}^{T}=\left( 
\begin{array}{ccc}
0 & a_{2} & a_{2}
\end{array}
\right) ,$ $\overline{\phi }_{3}=\left( 
\begin{array}{ccc}
0 & 0 & M
\end{array}
\right) $ and $\overline{\phi }_{23}=\left( 
\begin{array}{ccc}
0 & a_{2} & -a_{2}
\end{array}
\right) .$ In \cite{kingross} it was shown how such vacuum alignment can be
achieved in a supersymmetric theory, with $M>a_{2}$.

With this pattern of symmetry breaking it is easy to generate the required
form of the mass matrices via the superpotential 
\begin{eqnarray}
P_{Yukawa} &=&(\psi _{i}^{c}\phi _{3}^{i}\phi _{3}^{j}\psi _{j}+\psi
_{i}^{c}\phi _{23}^{i}\phi _{23}^{j}\psi _{j})H_{\alpha }/M^{2}+\epsilon
^{2}\epsilon ^{ijk}\psi _{i}^{c}\overline{\phi }_{23,j}^{i}\psi
_{k}^{c}H_{\alpha }/M  \label{yukawa} \\
&&+\epsilon ^{6}\left( \epsilon ^{ijk}\psi _{i}^{c}\overline{\phi }%
_{23,j}^{i}\overline{\phi }_{3,k}^{c}\right) \left( \epsilon ^{ijk}\psi _{i}%
\overline{\phi }_{23,j}^{i}\overline{\phi }_{3,k}^{c}\right) H_{\alpha
}/M^{4}  \nonumber
\end{eqnarray}
where $H_{1,2}$ are the Higgs doublets of the MSSM, which are $SU(3)$
singlets. If $\psi ^{c}=$ $u_{i}^{c},$ $d_{i}^{c}$ then $\alpha =2,$ $1$
respectively, giving rise to the up and down quark masses in a
supersymmetric theory. In eq(\ref{yukawa}) we have suppressed the couplings
of $O(1)$ associated with each operator. The first two operators shown are
the leading (dimension 6) ones consistent with a simple family symmetry%
\footnote{%
In a specific realisation of the scheme it is necessary to introduce further
Abelian family symmetries to ensure this is the most general form. A
specific realisation is given in reference \cite{kingross}.}\cite{kingross}.
The third and fourth operators are only generated via higher dimension
operators involving additional powers of fields $\phi _{23},$ $\phi _{3.}$
Replacing these fields by their vevs leads to the suppression factors $%
\epsilon ^{2}$ and $\epsilon ^{6}$ shown, where $\epsilon =a_{2}/M.$ This
gives rise to the mass matrices of the form 
\begin{equation}
\frac{M}{M_{3,3}}=\left( 
\begin{array}{ccc}
\lambda ^{\prime }\epsilon ^{8} & \lambda \epsilon ^{3} & \lambda \epsilon
^{3} \\ 
-\lambda \epsilon ^{3} & \lambda ^{\prime \prime }\epsilon ^{2} & \lambda
^{\prime \prime }\epsilon ^{2} \\ 
-\lambda \epsilon ^{3} & \lambda ^{\prime \prime }\epsilon ^{2} & 1+\lambda
^{\prime \prime }\epsilon ^{2}
\end{array}
\right)  \label{nonabelianmatrix}
\end{equation}
where the expansion parameter $\epsilon $ may differ for the up and down
sectors and we have explicitly included the $O(1)$ coefficients $\lambda ,$ $%
\lambda ^{\prime },$ $\lambda ^{\prime \prime }$ that arise due to the $O(1)$
couplings in eq(\ref{yukawa}) which are not related by the non-Abelian
symmetry. The effect of the $SU(3)$ family symmetry may be readily seen
relating the $(1,2)$ and $(1,3)$ matrix elements as well as the $(2,2)$ and $%
(2,3)$ matrix elements. Of course there are corrections to these equalities
coming at higher order in $\epsilon $ from operators of higher dimension. We
shall consider their effect in detail later.

The form of eq(\ref{nonabelianmatrix}) gives good agreement with the
measured quark mass matrices and can also be extended to give a viable
description of charged lepton and neutrino masses and mixing angles \cite
{kingross}. Here we consider whether the $O(1)$ coefficients can be related
by further symmetries and whether it is possible to determine the lepton
masses and mixing angles in terms of the quark masses and mixing angles.

\subsection{$SU(2)_{R}$ and Grand Unification}

In implementing the $SU(3)$ family symmetry the family structure has been
strongly constrained to generate left-right symmetry by assigning the left-
and the charge conjugate of the right- handed components of a given quark to
the same representation. Since, in addition, the family symmetry must
respect the $SU(2)_{L}$ gauge symmetry the up and down left-handed
components of a given quark family doublet must also carry the same family
charge. As a result of these two constraints one immediately sees that left-
and charge conjugate right- handed components of the up and down quark
members of a given family have the same family charge. This means that the
up and down quark mass matrices have the same {\it form} as in eqs(\ref{mu})
and (\ref{md}), although the expansion parameters may be different and the
operator coefficients of $O(1)$ may differ.

If $SU(2)_{R}$ is also an exact symmetry of the theory even the expansion
parameters and operator coefficients are equal and the mass matrices are
identical. This is clearly not acceptable and, if there is an underlying $%
SU(2)_{R}$ symmetry, it must be spontaneously broken so that the equality of
the mass matrices will be lost through soft symmetry breaking terms. These
enter through the expansion parameter $\epsilon $ which is determined by $%
\theta /M$ where $\theta $ is the field spontaneously breaking the symmetry
and $M$ is the messenger mass of the state responsible for communicating the
symmetry breaking and generating the higher dimension operators. Due to $%
SU(2)_{R}$ breaking the messenger mass may be different for the up and down
quark sectors and hence the expansion parameters may differ. It is also
possible that the family symmetry breaking field, $\theta ,$ is not a
singlet under $SU(2)_{R}$ and its vev breaks $SU(2)_{R},$ again leading to a
different expansion parameter for the up and the down sectors. However there
will still be some measurable effects of an underlying structure $SU(2)_{R}$
in the quark masses because the Yukawa couplings of the field $\theta $ to
the quarks and the messenger sector, which are responsible for the $O(1)$ \
coefficients, will respect the symmetry. How large are the corrections to
this result? They occur in radiative order but, in a supersymmetric theory,
supersymmetry guarantees that superpotential is not renormalised. In this
case the dominant radiative corrections to the Yukawa couplings come from
the D-terms controlling wavefunction renormalisation. Thus we may expect
that fields falling into $SU(2)_{R}$ doublets will have normalisation which
differ only in higher order, $O(|\theta |^{2}/M^{2}).$ The resulting
corrections to the $O(1)$ coefficients will also be at this order.

What are the phenomenological implications of an underlying $SU(2)_{R}?$ If
only the leading operator contributions to eq(\ref{nonabelianmatrix}) are
included the coefficients $a,$ $b$ and $c$ in eq(\ref{md}) will equal $%
a^{\prime },$ $b^{\prime }$ and $c^{\prime }$ in eq(\ref{mu}) respectively%
\footnote{%
In general we expect higher order operators contributing to a given matrix
element at $O(\theta /M)$ or above relative to the leading term.}. We wish
to explore whether such a broken symmetry expansion is consistent with the
quark and lepton masses and mixings. Unfortunately the tests of the $%
SU(2)_{R}$ relations in the quark sector are somewhat imprecise at present
because the contribution to the mixing angles from the up sector is small
and sensitive to small changes in the down sector. As a result $a^{\prime }$
and $c^{\prime }$ are only poorly determined. Given that in eqs(\ref{mu})
and (\ref{md}) the $(2,2)$ matrix elements have been used to define the
expansion parameters, the remaining test is to compare $b$ and $b^{\prime }.$
It is easy to eliminate the dependence on the expansion parameters by
combining $(1,1),$ $(2,2)$ and $(1,3)$ elements. Using eqs(\ref{mu}) and (%
\ref{md}) this gives 
\begin{equation}
\frac{b^{\prime }}{b}=\frac{m_{s}}{m_{c}}\sqrt{\frac{m_{u}m_{t}}{m_{d}m_{b}}}
\label{su2r}
\end{equation}
where this relation applies at the unification scale where $SU(2)_{R}$ is a
good symmetry. This gives 
\begin{equation}
\frac{b^{\prime }}{b}=0.5\pm 0.3  \label{su2rtest}
\end{equation}
Given that the expansion parameter in the down quark sector is quite large, $%
\overline{\epsilon }=0.15,$ and that higher order corrections may be
significant, the result of eq(\ref{su2rtest}) may be consistent with an
underlying $SU(2)_{R}$ relation prediction between $b^{\prime }$ and $b$,
with a significant correction at $O(\theta /M)$ coming from such higher
dimension operators. We shall return to a discussion of this possibility in
Section \ref{fit}.

Of course it is likely that $SU(2)_{R}$ is part of an underlying GUT such as 
$SO(10).$ In this case the question whether $SU(2)_{R}$ is an
approximate symmetry of the couplings depends on the pattern of symmetry
breaking. We shall also discuss a second possibility for reconciling the GUT
prediction with eq(\ref{su2rtest}) in which the couplings themselves feel $%
SU(2)_{R}$ breaking.

\subsection{GUTs and charged lepton masses}

We have seen that the quark mass matrices are consistent with the choice
that all the states of the (left-handed) components of a given family have
the same transformation properties under the family symmetry. This is
suggestive of a larger underlying GUT symmetry. The GUT $SO(10)$ is
particularly promising as all the (left-handed) states of a family, plus the
charge conjugate of the right handed neutrino, fit into a single $16$
dimensional representation. If $SO(10)$ is an underlying symmetry of the
theory any family symmetry must commute with it implying all the charges of
a given family must be the same, consistent with the form of the mass matrix
discussed above. An associated advantage of such a family symmetry is that
the mixed anomalies of the Standard Model gauge group with the family
symmetry will automatically cancel because of the structure of the
underlying $SO(10).$ It also contains $SU(2)_{R}$ and can relate the up to
the down sectors as discussed above. In addition a GUT symmetry can relate
quark and lepton mass matrices and this is the issue we wish to study here.

We first consider the charged leptons. Given the same family properties the
form of the charged lepton mass matrix will be the same as that of the down
quark matrices\ in eq(\ref{md}) up to coefficients of $O(1).$ If there is an
underlying GUT the coefficients too may be related. As mentioned above the
relation $m_{b}=m_{\tau }$ at the unification scale is in good agreement
with the measured masses. Such an equality applies in $SU(5)$ if the Higgs
responsible for the third generation masses transforms as a $\overline{5}$
of $SU(5).$ In $SO(10)$ equality applies if the Higgs belongs to a $10$  representation, this also gives equality between
the top quark and the third generation Dirac neutrino mass, something we
explore below.

What about the two lighter generations? In this case we must address the
question whether the expansion parameters are related. From a
phenomenological point of view, note that, after taking radiative
corrections into account, the relation $Det[M^{d}]=Det[M^{l}]$ at the
unification scale is also in good agreement with the experimental
measurements. From eq(\ref{nonabelianmatrix}) we see $Det[M]/M_{33}=\lambda
^{2}\epsilon ^{6}$ and so equality requires that the $(1,2)$ matrix element
of magnitude $\lambda \epsilon ^{3}$ be the same for the down quarks and the
leptons. This is consistent with an underlying broken $SU(2)_{R}$ symmetry
because the down quarks and leptons are both $T_{R,3}=-1/2$ states and both
can acquire their mass from the same Higgs doublet, $H_{2},$ in a
supersymmetric theory.\ Thus the strong $SU(2)_{R}$ breaking, needed to
split the $T_{R,3}=\pm 1/2$ states, is consistent with this equality.

\subsubsection{Symmetry breaking expansion parameters\label{exp}}

Of course equality of the down quark and charged lepton matrix elements in the $(1,2)$ position requires that the
expansion parameters be the same in the two sectors. This is consistent with
an underlying broken $SU(2)_{R}$ symmetry because the down quarks and
leptons are both $T_{R,3}=-1/2$ states and both can acquire their mass from
the same Higgs doublet, $H_{2},$ in a supersymmetric theory.
 Thus the strong $SU(2)_{R}$ breaking, needed to split the $T_{R,3}=\pm 1/2$ states,
does not differentiate between the down quarks and charged leptons. Clearly
if the dominant messenger sector for family symmetry breaking is in the
Higgs sector the expansion parameter in these two sectors will be equal.
Similarly, in this case, the expansion parameter in the up quark and
neutrino sectors will be the same, since both acquire Dirac masses from the
same Higgs doublet, $H_{1}.$ A similar conclusion applies if the family
symmetry breaking field, $\theta ,$ is not a singlet under $SU(2)_{R}$ and
its vev breaks $SU(2)_{R},$ again leading to a different expansion parameter
for the up and the down sectors.
        
The other possibility is that there are
significant contributions from messengers carrying quark and lepton quantum
numbers. For the messengers carrying left-handed generation quantum numbers,
$SU(2)_{L}$ requires the messenger masses should be equal so the only way
that these terms could be consistent with the up and down masses is if the
family symmetry breaking field, $\theta ,$ is not a singlet under
$SU(2)_{R}$ and its vev breaks $SU(2)_{R}.$ Finally it is possible that the
messengers carry the quantum numbers of the left-handed charge congugate
generations. In this case the up and down sectors could, through $SU(2)_{R}$
breaking, have a different masses leading to different expansion parameters.
If the underlying symmetry breaking pattern is
\begin{equation}
SO(10)\rightarrow SU(4)\times SU(2)_{L}\times
SU(2)_{R}\rightarrow SU(4)\times SU(2)_{L}\times U(1)
\end{equation}
the down quarks and charged leptons will have the same expansion parameter and so will the up quarks and neutrinos.

In what follows we will explore the
most predictive possibility, namely that there is an expansion paramete
$\overline{\epsilon }$ which applies to the down quarks and charged leptons,
and an expansion parameter $\epsilon $ which applies to the up quarks and
neutrinos.

\subsubsection{Quark/lepton couplings\label{couplings}}

If the expansion parameter is indeed the same for the down quarks and
leptons the quantities $Det[M^{d}]$ and $Det[M^{l}]$ occur at the same
order as is requires phenomenologically. Exact equality requires also the
equality of the coefficients determining the $(1,2)$ matrix elements. Just
as for the $(3,3)$ elements such equality will follow if the Higgs
responsible for this element transforms as a $\overline{5}$ of $SU(5)$ or to
a $10$ dimensional representation of $SO(10).$ However it is not possible to
have {\it identical} charged lepton and down quark mass matrices because,
after taking account of the radiative correction on going from high to low
scale, giving approximately a factor of 3 increase in the quark masses, the
relations $m_{s}=m_{\mu }$ and $m_{d}=m_{e}$ are in gross disagreement with
experiment. As pointed out by Georgi and Jarlskog \cite{Georgi:1979df} this discrepancy is
readily explained if there is an underlying GUT through the appearance of
Clebsch Gordon factors in the matrix element coefficients. In particular if
the Higgs responsible for the $(2,2)$ matrix element should belong to a $%
\overline{45}$ of $SU(5)$ (or $126$ of $SO(10)$) the lepton coupling is a
factor $-3$ times the down quark coupling. In this case, taking account of
the equality of the determinants, the relations for the light generations
are modified to give $m_{s}=m_{\mu }/3$ and $m_{d}=3m_{e}.$ Including the
radiative corrections needed to determine the masses at laboratory scales,
these relations are in excellent agreement with the measured masses. If
there is also an underlying non-Abelian family symmetry relating the matrix
elements as in eq(\ref{nonabelian}), the $(2,2)$ and $(2,3)$ matrix elements
are related and so it is necessary that the Higgs field responsible for the $%
(2,3)$ matrix element be also due to a Higgs the $\overline{45}$ of $SU(5).$
With this the resulting form of the lepton mass matrix coming from eq(\ref
{md}) is given by 
\begin{equation}
M^{l}\simeq \left( 
\begin{array}{ccc}
\overline{\epsilon }^{8} & b\overline{\epsilon }^{3} & c\overline{\epsilon }%
^{3} \\ 
b\overline{\epsilon }^{3} & -3\overline{\epsilon }^{2} & -3a\overline{%
\epsilon }^{2} \\ 
c\overline{\epsilon }^{3} & -3a\overline{\epsilon }^{2} & 1
\end{array}
\right)  \label{leptonmass}
\end{equation}
Note that in order to realise this scheme there must be a very particular
origin to the effective Lagrangian given in eq(\ref{yukawa}). The effective
operators may be generated through quark or Higgs mixing with states
carrying different family symmetry quantum numbers. In order to generate the
factor $3$ it is necessary that the operators responsible for these terms be
dominantly given by Higgs mixing of the $\overline{5}$ Higgs responsible for
the $(3,3)$ element with the $\overline{45}$ generating the factor $-3$
enhanced terms. Note that it is not necessary for the $\overline{45}$ to be
an elementary Higgs field, it can arise as an effective Higgs, for example
from the coupling of a $\overline{5}$ and $24$ (in $SO(10)$ the $126$ can be
generated by a $10$ and $45$) as in Figure 1\cite{raby}. In
these graphs $\overline{\chi }$ and $\chi $ ($\overline{\chi ^{c}}$ and $%
\chi ^{c}$) are a vectorlike pair of chiral supermultiplets with mass $M$ ($%
M^{\prime }),$ where $\chi $ ($\chi ^{c})$ has the same SM quantum numbers
as the left-handed generations $\psi $ (left-handed antigenerations $\psi
^{c}).$

Since the adjoint fields are
already needed to break $SU(5)$ (or $SO(10)$) this represents a
simplification of the Higgs sector. Moreover we see that such graphs can
also generate a coupling to an effective $120$ dimensional representation of 
$SO(10)$. This coupling would be antisymmetric in family space if the $120$
were fundamental but need not be so in the case they are given by Figure 1 because the intermediate messenger masses associated with the
graphs coupling to $H_{10}$ and $\Sigma _{45}$ are not necessarily the same
as those coupling to $\Sigma _{45}$ and $H_{10}.$ Writing the vacuum
expectation value
\begin{equation}
<\Sigma >=B-L+\kappa T_{R,3},  \label{effectivehiggs}
\end{equation}
\begin{figure}[ht*] 
\label{treegraph}
\centering
\mbox{\epsfig{file=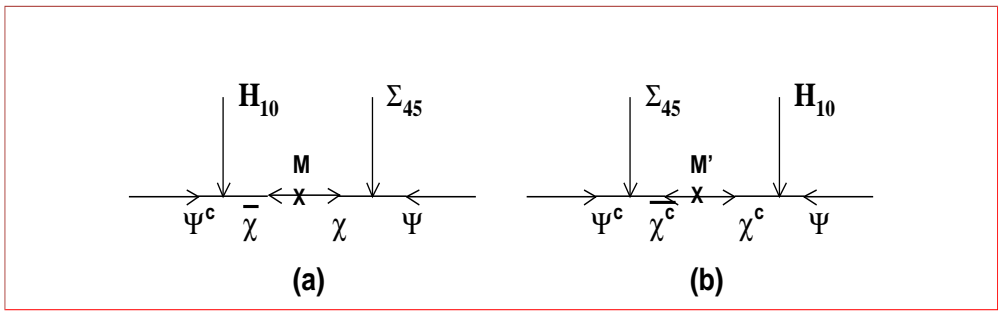, height=1.6in, width=5.2in}}
\caption{Froggatt Nielsen supergraphs generating fermion masses.}
\end{figure}
the relative contribution of these graphs to the down quarks and leptons is
respectively $1/3$ and -$1$ for Figure 1a and $(1/3-\kappa /2)$
and $(-1-\kappa /2)$ for Figure 1b. Consider the case the
second graph has lighter messengers and dominates. For $\kappa =0$ we obtain
the form of eq(\ref{leptonmass}). However for case $\kappa =2$ the
contribution to the leptons is $+3$ times that of the quarks so we obtain eq(\ref{leptonmass}) with $+3$ rather than $-3.$ Since the lepton mass
eigenvalues are insensitive to this sign we obtain identical masses for
either form. This case corresponds to the coupling of an effective $120$
dimensional representation and has different implications for the structure
of the up quarks and neutrinos that we explore below.

The mass matrix of eq(\ref{leptonmass}) gives excellent relation between the
down quarks and charged lepton masses of all three generations. Due to the
approximate texture zero in the $(1,1)$ position it also implies a
contribution to the $(1,2)$ element of the matrix $V^l$ diagonalising the charged lepton mass matrix, given approximately by
\begin{equation}
V^{l}_{12}\simeq \sqrt{\frac{m_{e}}{m_{\mu }}}  \label{nuemix}.
\end{equation}
In turn this gives a contribution to the  $(1,2)$ element of the MNS mixing matrix, $V^{MNS}=V^{l\dagger}V^{\nu}$ \cite{Maki:1962mu}. As we shall discuss this is significant in solar neutrino oscillations if the
(now disfavoured) SMA solution applies; it is also important in determining $V^{MNS}_{13}$. On the other hand, due to the
smallness of the $(1,3)$ matrix element, the contribution of the $(2,3)$ element of $V^l$ to the MNS mixing matrix is small 
\begin{equation}
V_{23}^{l}\simeq \frac{m_{\mu }}{m_{\tau }}.
\end{equation}
This is the analogue to the relation in the down quark sector $V_{cb}\simeq
m_{s}/m_{b},$ which is in good agreement with the small value observed.
However it is too small to explain the large mixing angle observed in
atmospheric neutrino oscillation. This discrepancy is at the heart of the
apparent conflict between quark and lepton masses and we turn now to the
discussion how this conflict may be resolved.

\section{Neutrino masses.}

\subsection{Dirac mass}

Following the discussion in Section \ref{family} we immediately see that the 
$SU(3)$ family symmetry properties of the left- and right- handed neutrinos
must be the same as those of the charged leptons. As a result the form of
the neutrino Dirac mass matrix between the doublet neutrinos and the singlet
(right-handed) neutrinos should be the same as eq(\ref{leptonmass}),
although the expansion parameter may be different. Given the success of GUT
relations in the charged mass matrix it is obviously reasonable to explore
the possibility that neutrino masses are similarly related to quark masses.
As discussed above  the expansion parameter for neutrino
masses may be equal to that for up quark masses. This is consistent with an underlying $%
SU(2)_{R}$ (contained in $SO(10))$ because the up quarks and right-handed
neutrinos are both $T_{3,R}=1/2$ states. In this case the resulting Dirac
neutrino mass matrix is of the form 
\begin{equation}
M_{D}^{\nu }/M_{D,33}^{\nu }=\left( 
\begin{array}{ccc}
O(\epsilon ^{8}) & b_{\nu }^{\prime }\epsilon ^{3} & c_{\nu }^{\prime
}\epsilon ^{3} \\ 
b_{\nu }^{\prime }\epsilon ^{3} & d_{\nu }^{\prime }\epsilon ^{2} & a_{\nu
}^{\prime }\epsilon ^{2} \\ 
c_{\nu }^{\prime }\epsilon ^{3} & a_{\nu }^{\prime }\epsilon ^{2} & 1
\end{array}
\right)  \label{diracneutrino}
\end{equation}
The underlying GUT symmetry relates the renormalisable couplings of the
messenger states. For the case the Higgs responsible for the $(3,3),$ $(1,2)$
and $(2,1)$ matrix elements transforms as the $10$ of $SO(10)$ and the
Higgs responsible for the $(2,2),$ $(2,3)$ and $(3,2)$ matrix elements
transforms as the $126$ of $SO(10)$ we have $a_{\nu }^{\prime }=-3a,$ $d_{\nu
}^{\prime }=-3,$ $b_{\nu }^{\prime }=b,$ and $c_{\nu }^{\prime }=c.$ In the
case just discussed where the Higgs responsible for the $(2,2),$ $(2,3)$ and 
$(3,2)$ matrix elements transform as the $120$ of $SO(10)$ (with the second 
graph of Figure 1 dominating) we have $a_{\nu }^{\prime }=-%
\frac{(6-3\kappa )}{(2+3\kappa )}a,$ $d_{\nu }^{\prime }=-\frac{(6-3\kappa )%
}{(2+3\kappa )},$ $b_{\nu }^{\prime }=b,$ and $c_{\nu }^{\prime }=c$

\subsection{Majorana mass}

Of course it is necessary to determine the Majorana mass matrix before one
can determine the effective neutrino mass matrix via the see-saw formula of
eq(\ref{seesaw}). Although the family symmetry properties of the
right-handed neutrinos are related to those of the charged leptons it is not
possible to use this information unambiguously to determine the structure of
the Majorana mass matrix. In particular it may not have the same form as is
found for the Dirac matrix. The reason is twofold. Firstly the Majorana
masses are generated via a new $\Delta L=2$ lepton number violating Higgs
sector and it is necessary to specify the family symmetry representation
content of this sector before the Majorana mass structure is fixed. Secondly
the Majorana mass matrix involves the coupling of identical fermions and so
antisymmetric terms allowed in the Dirac mass matrix will not arise in the
Majorana matrix. Despite this we can make some general statements about the
structure. In particular we expect an hierarchical structure for the
Majorana mass matrix because the underlying family symmetry ($SU(3)),$ is
the same as applies to the Dirac matrix which leads to a structure ordered
by a new expansion parameter $\varepsilon _{M}$.
 If a single $\Delta L=2$ (effective) symmetry breaking
field, $\Phi$  with definite $SU(3)$ family transformation properties, dominates there will no possibility of degeneracy between
matrix elements. For a large part of the parameter space, the form
of the Majorana mass matrix is in practice determined. In the limit that\ $%
\varepsilon _{M}<<\epsilon $ (which is the case if the messenger sector in
the $\Phi $ sector is heavier than in the electroweak breaking sector) the
mixing in the neutrino sector is dominated by the mixing coming from the
Dirac mass. If $\Phi$ couples dominantly to the $(3,3)$ position the resulting mass matrix can be diagonalised by
small rotations of $O(\varepsilon _{M})$ giving 
\begin{equation}
M_{\nu }=\left( 
\begin{array}{ccc}
m_{1} &  &  \\ 
& m_{2} &  \\ 
&  & m_{3}
\end{array}
\right)  \label{majneutrino1}
\end{equation}
It is important to note that this mass eigenstate basis is likely to be very
close to that used for the Dirac matrix, eq(\ref{diracneutrino}), the mass
eigenstates will be the family symmetry eigenstates up to corrections of $%
O(\varepsilon _{M}).$

This is the general form we shall use in our phenomenological analysis. It
corresponds to one of the two most reasonable possibilities. By
``reasonable'', we mean that large mixing should not be due to a detailed
correlation between the Dirac and Majorana mass matrix elements because this
would require a correlation between the $\Delta L=2$ and $\Delta L=0$
symmetry breaking sectors and there is no obvious symmetry that could
achieve this. The other reasonable possibility is that the mixing is
dominated by the mixing in the Majorana mass matrix (the case $\varepsilon
_{M}\gg \epsilon ).$This is less predictive (and hence less interesting)
because, in this case, we cannot relate neutrino mixing angles with quark
mixing angles. For this reason we concentrate here on the first possibility.

\section{The light neutrino mass matrix}

We are now able to determine the masses and mixing angles of the light
neutrinos using eqs(\ref{diracneutrino},\ref{majneutrino1}) in eq(\ref
{seesaw}). The effective Lagrangian associated with the see-saw mass is of
the form 
\begin{equation}
{\cal L}=\frac{\epsilon ^{6}}{m_{1}}(b_{\nu }^{\prime }\nu _{\mu }+c_{\nu
}^{\prime }\nu _{\tau }+O(\epsilon ^{5})\nu _{e})^{2}H_{1}^{2}+\frac{%
\epsilon ^{4}}{m_{2}}(d_{\nu }^{\prime }\nu _{\mu }+a_{\nu }^{\prime }\nu
_{\tau }+\epsilon b_{\nu }^{\prime }\nu _{e})^{2}H_{1}^{2}+\frac{1}{m_{3}}%
(\nu _{\tau }+a_{\nu }^{\prime }\nu _{\nu }+\epsilon c_{\nu }^{\prime }\nu
_{e})^{2}H_{1}^{2}  \label{effmass}
\end{equation}

\subsection{Near maximal mixing\label{maximal}}

It is now straightforward to determine the viable forms of the mass matrix.
Consider first the heaviest neutrino which should have near maximal mixing
to explain atmospheric neutrino oscillation. One sees from eq(\ref{effmass})
that, surprisingly, this is easy to achieve.

We first consider the case $\frac{\epsilon ^{4}}{m_{2}}>\frac{\epsilon ^{6}}{%
m_{1}},$ $\frac{1}{m_{3}}.$ The heaviest eigenstate is $\nu _{1}\propto
d_{\nu }^{\prime }\nu _{\mu }+a_{\nu }^{\prime }\nu _{\tau }+\epsilon b_{\nu
}^{\prime }\nu _{e}.$ If $SU(2)_{R}$ is an underlying symmetry of the
couplings ($\kappa =0$ in eq(\ref{effectivehiggs})), $a_{\nu }^{\prime
}/d_{\nu }^{\prime }\simeq a\simeq 1.3.$ In this case $\nu _{1}$ is very
close to a maximal mixed state of muon and tau neutrinos. Note, as remarked
in Section \ref{family}, it is the smallness of $V_{cb}$ which requires a $%
(2,3)$ entry of $O(\overline{\epsilon }^{2})$ that leads to near maximal
mixing in the neutrino sector\cite{kingross}!

It is also straightforward to compute the next lightest neutrino state. It
will have mass of $O(\frac{\epsilon ^{6}}{m_{1}}$ $or$ $\frac{1}{m_{3}})$
whichever is the larger. Let us consider the first possibility first. The
eigenstate $\nu _{2}$ has the form 
\begin{equation}
\nu _{2}\varpropto (a_{\nu }^{\prime }b_{\nu }^{\prime }-c_{\nu }^{\prime
}d_{\nu }^{\prime })(d_{\nu }^{\prime }\nu _{\tau }-a_{\nu }^{\prime }\nu
_{\mu })+O(\epsilon )\nu _{e}  \label{second state}
\end{equation}
Except for the case the factor $(a_{\nu }^{\prime }b_{\nu }^{\prime }-c_{\nu
}^{\prime }d_{\nu }^{\prime })$ is anomalously small the second eigenstate
will be predominately in the $\nu _{\mu },$ $\nu _{\tau }$ direction. The
same conclusion applies if the second eigenstate is dominated by the third
term of eq(\ref{effmass}). Thus in both cases the mixing angle $V_{\nu
_{e}\mu }$ relevant to solar neutrino oscillation will be dominated by the
contribution of $O(\overline{\epsilon })$ from the charged lepton sector
given in eq(\ref{nuemix}). This is of the correct magnitude to fit the SMA
solution. The ratio of the two heaviest masses is given by $%
M_{3}/M_{2}=O(\epsilon ^{2}m_{2}/m_{1}\;or\;\epsilon ^{4}m_{2}/m_{3})$ and
the Majorana masses can be adjusted to get the value needed for the SMA
solution. For the case $\frac{\epsilon ^{6}}{m_{1}}>\frac{\epsilon ^{4}}{%
m_{2}},\frac{1}{m_{3}}$ a similar pattern emerges except that the heaviest
state is now $\nu _{1}\propto b_{\nu }^{\prime }\nu _{\mu }+c_{\nu }^{\prime
}\nu _{\tau }.$ $SU(2)_{R}$ symmetry leads to the relation $b_{\nu }^{\prime
}/c_{\nu }^{\prime }\simeq b/c\simeq 0.5$ and so the mixing is large but not
maximal. Provided the factor $(a_{\nu }^{\prime }b_{\nu }^{\prime }-c_{\nu
}^{\prime }d_{\nu }^{\prime })$ is not anomalously small the lighter states
are not strongly mixed and the dominant contribution to $V^{MNS}_{12}$
comes from the charged lepton sector. From this we see that a large
atmospheric neutrino mixing angle and the SMA solar neutrino solution is
consistent with a very symmetrical structure for the mass matrices in which
there is an approximate up-down $SU(2)_{L}\times SU(2)_{R}$ symmetry
together with a GUT symmetry relating quarks and leptons.

This analysis has demonstrated that, through the see-saw mechanism, it is
quite natural to get near maximal mixing in the neutrino channel for
atmospheric neutrino mixing, the large lepton mixing being due to the small
mixing in the quark sector! However inclusion of the recent SNO data now
strongly disfavours the SMA solution for the solar mixing angle and favours
large mixing in this case too. To address this we now turn to the question
whether near bi-maximal mixing is consistent with an enlarged family and
flavour symmetry.

\subsection{Near bi-maximal mixing}

At first sight is seems that our analysis has already ruled out the
possibility of near bi-maximal mixing. The possible exception to this is if
the factor $(a_{\nu }^{\prime }b_{\nu }^{\prime }-c_{\nu }^{\prime }d_{\nu
}^{\prime })$ is of $O(\epsilon ),$ in which case the second state of eq(\ref
{second state}) has a large $\nu _{e}$ component. There are two ways that
this may happen naturally. The first is through $SU(2)_{R}$ breaking in the
coefficients which occurs through spontaneous breaking via the vev of eq(\ref
{effectivehiggs}) for the case $\kappa \neq 0$. 
For the special case $\kappa=2,$ which generates good charged lepton masses, we have $a_{\nu }^{\prime}=d_{\nu }^{\prime }=0.$ In this case higher order terms in the expansion in
terms of the symmetry breaking parameter can generate $(a_{\nu }^{\prime
}b_{\nu }^{\prime }-c_{\nu }^{\prime }d_{\nu }^{\prime })$ of $O(\epsilon )$
as required. The case $\kappa =2$ can arise quite naturally on spontaneous \
symmetry breaking as it corresponds to an enhanced symmetry point at which
the effective potential can easily have a minimum. The second way $(a_{\nu
}^{\prime }b_{\nu }^{\prime }-c_{\nu }^{\prime }d_{\nu }^{\prime })$ may
naturally be of $O(\epsilon )$ applies even to the case $\kappa =0$ and
again relies on higher order terms in the expansion in terms of the symmetry
breaking parameter.

Let us start with a discussion of the sub-leading terms. Consider the case of the $SU(3)$ family symmetry which leads to the leading
order mass matrix of the form given in eq(\ref{nonabelianmatrix}). Including
higher order terms in the symmetry breaking expansion parameter gives a mass
matrix of the form

\begin{equation}  \label{massmatpar}
M/M_{3,3}=\left( 
\begin{array}{ccc}
\varepsilon ^{8} & \varepsilon ^{3}(z+(x+y)\varepsilon) & \varepsilon ^{3}(z
+(x-y)\varepsilon ) \\ 
-\varepsilon ^{3}(z+(x+y)\varepsilon) & \varepsilon ^{2}(aw+u\varepsilon) & 
\varepsilon ^{2}(aw-u\varepsilon ) \\ 
-\varepsilon ^{3}(z +(x-y)\varepsilon) & \varepsilon ^{2}(aw-u\varepsilon) & 
1
\end{array}
\right)  \label{mdexp}
\end{equation}

Here $z$, $w$ and $u$ are real coefficients and $x$ and $y$ complex
coefficients of order 1. For the case of the down quarks and leptons the
expansion parameter is $\varepsilon =\bar{\epsilon}$ and for up quarks and neutrinos $%
\varepsilon =\epsilon $ ({\it cf.}the discussion of Section \ref{exp}). 
The higher order terms are necessary to fit the data for $M_{d}$ of eq(\ref
{md}). Following the discussion of Section \ref{couplings} we assume that
the coefficient $a$ arises from an effective $120$ or $126$\ $SO(10)$
representation in order to generate the correct lepton mass spectrum. Thus $%
a $ is determined by the value of $\kappa $ in eq(\ref{effectivehiggs}) and
there are two possibilities 
\begin{eqnarray}
\kappa &=&0,\;a_{l}=-3,\quad a_{\nu }=-3 \\
\kappa &=&2,\;a_{l}=+3,\quad a_{\nu }=0
\end{eqnarray}
We assume all other coefficients come from a $10$ of $SO(10)$ so they are
the same as for the quarks$.$

We may determine the neutrino effective mass matrix using the see-saw
formula, eq(\ref{seesaw}). We focus on the case that the dominant
contribution to the heaviest of the light neutrino states is due to the
exchange of the right handed neutrino with mass $m_{1}$ which couples to the
first row of the Dirac mass matrix (\ref{massmatpar}). As a result it is
given by

\begin{equation}
\nu _{a}=\frac{(z+x\epsilon )(\nu _{\mu }+\nu _{\tau })+y\epsilon (\nu _{\mu
}-\nu _{\tau })}{\sqrt{(z+(x+y)\epsilon )^{2}+(z+(x-y)\epsilon )^{2}}}
\end{equation}
The state orthogonal to it is 
\begin{equation}
\nu _{b}=\frac{(z+x\epsilon )(\nu _{\mu }-\nu _{\tau })-y\epsilon (\nu _{\mu
}+\nu _{\tau })}{\sqrt{(z+(x+y)\epsilon )^{2}+(z+(x-y)\epsilon )^{2}}}.
\end{equation}
We concentrate on the case that the dominant contribution to the second
heaviest state is due to the exchange of the right handed neutrino with mass 
$m_{2}$ which couples to the second row in the Dirac mass matrix. The mass
eigenstates are given by 
\begin{eqnarray}
\nu _{3} &\simeq &\nu _{a} \\
\nu _{2} &\simeq &\frac{|z|e^{i\xi }\nu _{e}+|r|\nu _{b}}{\sqrt{z^{2}+r^{2}}}
\\
\nu _{1} &\simeq &\frac{|r|\nu _{e}-|z|e^{-i\xi }\nu _{b}}{\sqrt{z^{2}+r^{2}}%
},
\end{eqnarray}
where 
\[
\xi =\phi _{z}-\phi _{r} 
\]
\begin{equation}
r=\frac{\sqrt{2}}{z}(zu-a_{\nu }wy)
\end{equation}
Note the importance of the higher order terms in determining the lighter
neutrino eigenstates. In particular, the $SU(3)$ family symmetry
aligns the leading terms in the $(2,2)$ and the $(2,3)$ directions and the $%
(1,2)$ and the $(1,3)$ directions; at this order $b^{\prime}_{\nu}=c^{\prime}_{\nu}=z$, $d^{\prime}_{\nu}=a^{\prime}_{\nu}=aw$ so the coefficient $a^{\prime}_{\nu}b^{\prime}_{\nu}-c^{\prime}_{\nu}d^{\prime}_{\nu}$ vanishes in leading order (this is the second case mentioned above). For this reason the second heaviest neutrino can have a
large $\nu _{e}$ component due to the higher order terms. Let us consider
the phenomenological implications of this fit. The mixing angle $\theta
_{23} $ relevant to atmospheric neutrino mixing is given by 
\begin{equation}
\tan ^{2}\theta _{23}\simeq \left| \frac{z+(x+y)\epsilon }{z+(x-y)\epsilon }%
\right| ^{2}.
\end{equation}
Note that it is relatively insensitive to the higher order terms of eq(\ref
{massmatpar}).

The mixing angle $\theta _{12}$ relevant to solar neutrino mixing is
given by 
\begin{equation}
\tan ^{2}\theta _{12}\simeq \left| \frac{z}{r}\right| ^{2}
\end{equation}
With these results we may now ask whether it is possible to obtain near
bi-maximal mixing with the parameterisation of eq(\ref{mdexp}), with the
parameters constrained to fit the quark masses. As the latter do not
determine all the parameters precisely, we will use the remaining freedom to
try to generate all the phenomenologically allowed cases, namely the LMA,
LOW and VAC solutions although we note that the recent SNO\ data prefers the
LMA solution.

\subsubsection{Solution for $\protect\kappa $=0}

\begin{table}[htbp] \centering%
%
\begin{tabular}{|c|c|}
\hline
$z$ & $0.91\pm 0.06$ \\ 
$x$ & $-(2.95\pm 0.1)$ \\ 
$y$ & $0.74\pm 0.07$ \\ 
$\phi _{x}$ & $0.05\pm 0.01$ \\ 
$\phi _{y}$ & $0.19\pm 0.01$ \\ 
$w$ & $0.57\pm 0.05$ \\ 
$u$ & $-(0.55\pm 0.12)$ \\ 
$\bar{\epsilon}$ & $0.21\pm 0.01$ \\ 
$\epsilon$ & $0.07\pm 0.01$\\
$t_{12}^{2}$ & $0.58\pm 0.02$ \\ 
$t_{23}^{2}$ & $1.49\pm 0.05$ \\ 
$\left( \frac{M_{2}}{M_{3}}\right) ^{2}<$ & ($0.8\pm 0.4)\times 10^{-5}$ \\ 
$a_{u}$ & $2.4\pm 0.3$ \\ \hline
\end{tabular}
\caption{The parameters for a fit to the down quark mass matrix under the
LOW constraint for $\theta_{12}$ in Table 6. \label{tlowfitafsno}}%
\end{table}%
%

\begin{table}[htbp] \centering%
%
\begin{tabular}{|c|c|}
\hline
$z$ & $0.95\pm 0.01$ \\ 
$x$ & $-(3.15\pm 0.31)$ \\ 
$y$ & $0.64\pm 0.03$ \\ 
$\phi _{x}$ & $0.039\pm 0.006$ \\ 
$\phi _{y}$ & $0.19\pm 0.01$ \\ 
$w$ & $0.53\pm 0.03$ \\ 
$u$ & $-(0.5\pm 0.1)$ \\ 
$\bar{\epsilon}$ & $0.22\pm 0.01$ \\ 
$\epsilon$ & $0.07\pm 0.01$\\
$t_{12}^{2}$ & $1.27\pm 0.04$ \\ 
$t_{23}^{2}$ & $1.4\pm 0.11$ \\ 
$\left( \frac{M_{2}}{M_{3}}\right) ^{2}<$ & ($6.0\pm 1.6)\times 10^{-6}$ \\ 
$a_{u}$ & $2.7\pm 0.1$ \\ \hline
\end{tabular}
\caption{The parameters for a fit to the down quark mass matrix under the
VAC constraint for $\theta_{12}$ in Table 6. \label{tvacfitafsno}}%
\end{table}%
%

For $\kappa =0$, i.e. $a_{\nu }=-3$, the condition assumed above that the
right handed neutrinos of mass $m_{1}$ and $m_{2}$ dominate the see-saw
contribution to the heaviest and next heaviest light neutrino masses
respectively is $\frac{z^2\epsilon ^{6}}{m_{1}}>\frac{9\omega ^{2}\epsilon ^{4}%
}{m_{2}}.$ In this case the masses of the light neutrino mass eigenstates
are given approximately by 
\begin{eqnarray}
M_{3} &=&\frac{\epsilon ^{6}}{m_{1}}2z^{2}v^{2} \\
M_{2} &=&\frac{\epsilon ^{6}}{m_{2}}z^{2}(1+\tan ^{-2}\theta _{12})v^{2} \\
M_{1} &<&\frac{1}{m_{3}}v^{2}
\end{eqnarray}
Thus for $\frac{z^2\epsilon ^{6}}{m_{1}}>\frac{9\omega ^{2}\epsilon ^{4}}{m_{2}}
$ we have 
\begin{equation}
\frac{M_{2}}{M_{3}}<\left( \frac{z}{3w}\right) ^{2}\left( \frac{1+\tan
^{-2}\theta _{12}}{2}\right) \epsilon ^{2}
\end{equation}
As we have discussed a large value for $\theta _{12}$ appears naturally.
However $M_{2}/M_{3}$ is of $O(\epsilon ^{2})$ and for choices of the
parameters $z,$ $\omega $ of $O(1)$ and $\theta _{12}$ in the observed
range, we can not obtain a large enough value for the ratio $M_{2}/M_{3}$ to
fit the LMA solution. Thus in this case we can only obtain the LOW and VAC
solutions of the data analysis of neutrino oscillations and mixings. The
remaining parameters are constrained by the requirement that the
parameterisation of equation (\ref{massmatpar}) fits the down quark mass
matrix (the first fit of eq(\ref{fit})). The results are presented in Tables 
\ref{tlowfitafsno} and \ref{tvacfitafsno}. These may be compared to the fits
to neutrino masses and mixing angles coming from the analysis of neutrino
oscillations; for convenience we summarise the present situation in the
Appendix in Tables \ref{appendixt1}, \ref{appendixt2},\ref{appendixt4} and 
\ref{appendixt5}. As may be seen we can obtain a good description of both the LOW and VAC solutions.

An important cross check of the structure of the GUT relations used in this
analysis comes from the up quark mass matrix. It is given by eq(\ref
{massmatpar}) with $\varepsilon =\epsilon $ and $a=a_{u}=1$ (since $\kappa
=0 $). The prediction $a_{u}=1$ is the analogue of the prediction $b^{\prime
}=b $ in eq(\ref{su2rtest}). Given the parameters $z$ and $\omega $ from the
fit of Tables \ref{tlowfitafsno} and \ref{tvacfitafsno}, we can test this
prediction by using the up quark masses to determine $a_{u}.$ The results
are also given in the Tables. We may see that the inclusion of the higher
order corrections has actually made the fit to the up quark mass matrix
worse than the case ($b^{\prime }=b$) tested in eq(\ref{su2rtest}) casting
doubt on the viability of the $\kappa =0$ solution.

\subsubsection{Solution for $\protect\kappa $=2}

\begin{table}[htbp] \centering%
%
\begin{tabular}{|c|c|}
\hline
$z$ & $0.78\pm 0.14$ \\ 
$x$ & $-(1.86\pm 0.51)$ \\ 
$y$ & $1.37\pm 0.21$ \\ 
$\phi _{x}$ & $0.14\pm 0.04$ \\ 
$\phi _{y}$ & $0.19\pm 0.05$ \\ 
$w$ & $0.77\pm 0.06$ \\ 
$u$ & $-(0.87\pm 0.15)$ \\ 
$\bar{\epsilon}$ & $0.18\pm 0.02$ \\ 
$\epsilon$ & $0.06\pm 0.01$\\
$t_{12}^{2}$ & $0.4\pm 0.1$ \\ 
$t_{23}^{2}$ & $1.7\pm 0.2$ \\ 
$\left( \frac{M_{2}}{M_{3}}\right) ^{2}$ & $(3\pm 0.11)$ $\left( \frac{m_{1}}{m_{2}}%
\right) ^{2}$ \\ 
$a_{u}$ & $1.6\pm 0.4$ \\ \hline
\end{tabular}
\caption{The parameters for a fit to the down quark mass matrix under the
LMA constraint for $\theta_{12}$ in Table 6. \label{lmasnopar}}%
\end{table}%
%

\begin{table}[tbp]
\label{lowkeq2}
\par
\begin{center}
\begin{tabular}{|c|c|}
\hline
$z$ & $0.95\pm 0.05$ \\ 
$x$ & $-2.80\pm 0.13$ \\ 
$y$ & $1.37\pm 0.16$ \\ 
$\phi_x$ & $0.09\pm 0.01$ \\ 
$\phi_y$ & $0.18\pm 0.01$ \\ 
$w$ & $0.77\pm 0.02$ \\ 
$u$ & $-0.86\pm 0.06$ \\ 
$t^2_{12}$ & $0.58\pm 0.04$ \\ 
$t^2_{23}$ & $1.75\pm 0.18$ \\ 
$\bar{\epsilon}$ & $0.19\pm 0.03$ \\ 
$\epsilon$ & $0.07\pm 0.01$\\
$\left(\frac{M_2}{M_3}\right)^2$ & $(1.77\pm 0.11)\left(\frac{m_1}{m_2}\right)^2$ \\ 
$a_u$ & $1.9\pm 0.1$ \\ \hline
\end{tabular}
\end{center}
\caption{{\protect\small {The parameters for a fit to the down quark mass
matrix for the solution $k=2$, under the LOW constraint for $\protect\theta %
_{12}$.}}}
\label{tlowfitafsnok2}
\end{table}

\begin{table}[tbp]
\label{vackeq2}
\par
\begin{center}
\begin{tabular}{|c|c|}
\hline
$z$ & $0.96\pm 0.06$ \\ 
$x$ & $-3.20\pm 0.13$ \\ 
$y$ & $0.87\pm 0.13$ \\ 
$\phi_x$ & $0.05\pm 0.01$ \\ 
$\phi_y$ & $0.19\pm 0.01$ \\ 
$w$ & $0.61\pm 0.02$ \\ 
$u$ & $-0.61\pm 0.04$ \\ 
$t^2_{12}$ & $1.23\pm 0.01$ \\ 
$t^2_{23}$ & $1.55\pm 0.20$ \\ 
$\bar{\epsilon}$ & $0.20\pm 0.03$ \\ 
$\epsilon$ & $0.07\pm 0.01$\\
$\left(\frac{M_2}{M_3}\right)^2$ & $(0.82\pm 0.1)\left(\frac{m_1}{m_2}\right)^2$ \\ 
$a_u$ & $2.4\pm 0.2$ \\ \hline
\end{tabular}
\end{center}
\caption{{\protect\small {The parameters for a fit to the down quark mass
matrix for the solution $k=2$ and under the VAC constraint for $\protect%
\theta _{12}$.}}}
\label{tvacfitafsnok2}
\end{table}

For $\kappa =2$, i.e. $a_{\nu }=0$, the condition that the right handed
neutrinos of mass $m_{1}$ and $m_{2}$ respectively dominate the see-saw
contribution to the heaviest and next heaviest light neutrino eigenstates
masses is $\frac{\epsilon ^{6}}{m_{1}}>O(\frac{\epsilon ^{6}}{m_{2}}).$ In
this case the masses of neutrino mass eigenstates are given approximately by 
\begin{eqnarray}
M_{3} &=&\frac{\epsilon ^{6}}{m_{1}}2z^{2}v^{2} \\
M_{2} &=&\frac{\epsilon ^{6}}{m_{2}}(2u^{2}+z^{2})v^{2} \\
M_{1} &<&\frac{1}{m_{3}}v^{2}
\end{eqnarray}

\begin{equation}
\frac{M_{2}}{M_{3}}\approx \frac{2u^{2}+z^{2}}{2z^{2}}\left( \frac{m_{1}}{%
m_{2}}\right) .
\end{equation}

Since the condition $\frac{z^2\epsilon ^{6}}{m_{1}}>O(\frac{\epsilon ^{6}}{%
m_{2}})$ is readily satisfied for $m_{1}/m_{2}\lesssim O(1)$ the LMA
solution for neutrinos can readily be reproduced through a choice of the
ratio $m_{1}/m_{2}$. The results of the fit to the down quark mass matrix is
given in Table \ref{lmasnopar}. This solution is obtained when $\tan \theta
_{12}^{2}=0.4$, which is the favoured value for the LMA solution. As may be
seen from a comparison with Tables \ref{appendixt1}, \ref{appendixt2} and 
\ref{appendixt3}, the agreement with neutrino mixing angles is good.

For the case $\kappa =2$ the prediction for the up quark mass matrix is that 
$a_{u}=2.$ We may see from Table \ref{lmasnopar} that this is in better
agreement with the up quark masses than the case $\kappa =0$ and also with
the case tested in tested in eq(\ref{su2rtest}) when the effects of higher
order operators were not included. This is an encouraging indication that the
case $\kappa =2$ may be realised.

It is also possible to obtain the LOW and VAC solutions for $\kappa =2.$ The
results are presented in Tables \ref{tlowfitafsnok2} and \ref{tvacfitafsnok2}%
. These may be compared to the fits to neutrino masses and mixing angles
coming from the analysis of neutrino oscillations in good agreement with the
fits to experiment which are summarised in the Appendix in Tables \ref
{appendixt1}, \ref{appendixt2},\ref{appendixt4} and \ref{appendixt5}. In
contrast to the $\kappa =0$ case, the value for $a_{u}$ is in excellent
agreement with the measured up quark masses.

\subsection{The prediction for $V_{13}$}

Our analyses has so far investigated the implications for $V_{23}$ and $%
V_{12}$ which follow from an enhanced flavour and family symmetry. What are
the implications for $V_{13}?$ This is a particularly interesting question
for only if $V_{13}$ is quite large will there be any prospect of seeing CP
violation in future long baseline neutrino experiments. Due to the form
of the hierarchy in the mass matrices, both the charged leptonic mass matrix
and the effective neutrino mass matrix, it is possible to diagonalise them
by the rotations $R_{23}R_{13}R_{12}$ together with diagonal matrices
carrying the phases. From this the $13$ ($e3)$ $V^{MNS}$ mixing matrix is
given by \cite{VeMNCP,King:2002nf} 
\begin{equation}
V^{MNS}_{13}=s_{13}^{\nu }c_{12}^{l}c_{13}^{l}e^{i\omega
_{1}}-s_{13}^{l}c_{12}^{l}c_{13}^{\nu }c_{23}^{\prime }e^{i\omega
_{2}}+s_{12}^{l}s_{23}^{L\prime }c_{13}^{\nu }e^{i\omega _{3}}
\label{eleUe3}
\end{equation}
where $s_{ij}^{f}$, for $i,j=1,2,3$ and $f=l,\nu $ represent the sines of
the mixing angles, analogously for the cosines and the phases $\omega _{i}$, 
$i=1,2,3$ are functions of the phases appearing in the mass matrix. The
angle $s_{23}^{L\prime }$ is given by $|c_{23}^{\nu }s_{23}^{l}-s_{23}^{\nu
}c_{23}^{l}e^{i\omega _{4}}|$. Since the neutrino oscillations experiments are not sensitive to Majorana CP phases, the standard parametrisation  used  for the case of quarks \cite{Hagiwara:2002pw} is commonly used for the case of leptons.
 In this parameterisation $%
V_{13}=s_{13}e^{-i\delta }$, where $s_{13}$ should be identified with the
sine of the CHOOZ angle and $\delta $ is the analogue to the quark CP  violation phase. Thus identifying $s_{13}$ with the absolute value
of eq(\ref{eleUe3}) we can trace the contributions from charged leptons and
neutrinos to the CHOOZ angle.

We can see that, in both the $\kappa =0$ and $\kappa =2$ cases discussed
above, the value of $s_{13}^{\nu }$ is negligible because of the hierarchy
of the Majorana masses we have taken ($m_{3}>>m_{2},m_{1}$). The mixings in
the leptonic sector are small and are given by 
\begin{equation}
s_{12}^{l}\approx \frac{|Y_{12}^{l}-\frac{Y_{13}^{l}Y_{23}^{l}}{Y_{33}^{l}}|%
}{|Y_{22}^{l}-\frac{(Y_{23}^{l})^2}{Y_{33}^{l}}|};\qquad s_{13}^{l}\approx \frac{%
|Y_{13}^{l}+\frac{Y_{12}^{l}Y_{23}^{l}}{Y_{33}^{l}}|}{|Y_{33}^{l}|},
\label{sin1213cle}
\end{equation}
where $Y_{12}^{l}$ are the elements of the charged lepton mass matrix. From eq(\ref{massmatpar}) we see that $s_{13}^{l}=O(\bar{\epsilon}%
^{3})$ and $s_{12}^{l}=O(\bar{\epsilon}^{2})$, therefore the latter is
the dominant contribution. This result is interesting because $%
Y_{12}^{l}/Y_{22}^{l}\approx \sqrt{m_{e}/m_{\mu }}$ and hence from eqs(\ref
{eleUe3}) and (\ref{sin1213cle}) 
\begin{equation}
V^{MNS}_{13}\simeq \sqrt{\frac{m_{e}}{m_{\mu }}}
\end{equation}
which is a testable prediction. This gives a value $V^{MNS}_{13}\approx 0.07,$ close to
the present bound and large enough for the future CP violating experiments
to be viable. This point has been made explicitly by King \cite{King:2002nf}
and the value given is consistent with the results of \cite{Lavignac:2002gf}\cite{Frigerio:2002rd}.
When the third term in equation (\ref{eleUe3}) is dominant, $\delta$, the analogue to the quark CP violation phase, is  $-\omega_3$, which in terms of the elements, $|Y^{\nu}_{ij}|e^{i\phi_{ij}}$, of the effective neutrino mass matrix it is given by:
\begin{equation}
\delta=-\omega_3\approx \frac{\gamma^{\nu}_{23} -\gamma^{\nu}_{12}-\gamma^{\nu}_{13}}{2}
\end{equation}
where 
\bea
\tan\gamma^{\nu}_{ij}&\approx& \frac{Y^{\nu}_{ij}Y^{\nu}_{jj}\sin(\phi_{ij}-\phi_{jj})+ Y^{\nu}_{ik}Y^{\nu}_{jk}\sin(\phi_{ik}-\phi_{jk})}{Y^{\nu}_{ij}Y^{\nu}_{jj}\cos(\phi_{ij}-\phi_{jj})+ Y^{\nu}_{ik}Y^{\nu}_{jk}\cos(\phi_{ik}-\phi_{jk})}\\
\gamma^{\nu}_{23}&\approx& \phi_{23}-\phi_{33}
\eea
for $ij=12,13$ and $k\neq i,j$.
In the same way as discussed for the quarks, Section \ref{quarkmassmats}, we may choose the phases to be $\phi^l$, the phase of the $(1,2)$ element of the Dirac $M^{\nu}$ and $\chi^l$ the phase of the $(1,3)$ element of $M^l$, then  $\delta\approx\phi^l$. 

The process of extracting the angles and phases from successive rotations is
general and can be applied directly to hierarchical mass matrices; for a
further work on this applied to leptons and the implications for CP
violation see \cite{VeMNCP}.

\subsubsection{\protect\bigskip $\protect\mu \rightarrow e\protect\gamma $}

Several groups \cite{Casas:2001sr,Lavignac:2001vp} have pointed out that off diagonal lepton Yukawa couplings can lead to unacceptably large contributions for lepton family number violation processes such as $\mu \rightarrow e\gamma$. In a supersymmetric framework the branching ratio, $BR(\mu \rightarrow e\gamma)$ has the form
\begin{equation}
\label{brmugamma}
BR(\mu \rightarrow e\gamma)\propto |(Y_{\nu }^{\dagger }\ln\frac{M_X}{M_R}Y_{\nu })_{21}|^{2} \tan ^{2}\beta,
\end{equation}
where $Y_{\nu }$ is the matrix of Yukawa couplings for Dirac neutrinos (in the basis in which charged leptons are flavour diagonal), $M_{X}$ is the scale of Grand Unification, $M_{R}$ is the scale where right handed neutrinos decouple and 
$\tan \beta $ is equal to the ratio of the vacuum expectation values of the
two Higgs doublets of the MSSM. Since $BR(\mu \rightarrow e\gamma)$ depends as well on the soft supersymmetric mass spectrum is useful to determine instead the matrix element $C_{\mu e}=(Y_{\nu }^{\dagger }\ln (M_{X}/M_{R})Y_{\nu })_{21}$  and hence study the predictions of the structure of $Y_{\nu}$  and the scale of $M_R$ for lepton flavour violating processes \cite{Lavignac:2001vp}.

 We can compute the element $C_{\mu e}$  for our
texture of $Y_{\nu }$. For the preferred case $\kappa =2$ we have $C_{\mu
e}\approx 6\times 10^{-3}$ for the LMA solution, $C_{\mu e}\approx 5.6\times 10^{-3}$
for the LOW solution and $C_{\mu e}\approx 4.2\times 10^{-3}$ for VAC solution .  Even for the case that $\tan \beta $ is
large ($h_{t}\simeq h_{b})$, due to the hierarchical form of the Yukawa
couplings, eq(\ref{massmatpar}), $|(Y_{\nu }^{\dagger }Y_{\nu })_{21}|\tan \beta$ is small and therefore does not conflict the bounds for $\mu \rightarrow e\gamma$ {\footnote {The structure of the effective mass matrix for low energy neutrinos (provided by the
see-saw mechanism) in the models considered here belongs to the Class 3 of
reference \cite{Lavignac:2001vp}.}}.
As can be seen from Figure 2 \cite{Lavignac:2001vp} the coefficients $C_{\mu e}$ given above  fall below current experimental bounds \cite{Brooks:1999pu} for a wide range of the soft supersymmetric breaking parameter space {\footnote {To compare with the bounds presented in \cite{Lavignac:2001vp} we need to multiply the upper bounds  for $C_{\mu e}$  by $10/50$, since  $\tan \beta \approx 50$.}}.

\subsection{Summary and conclusions}

Due to the see-saw mechanism the significant differences between quark and
lepton mixing can be explained while keeping the form of their Dirac mass
matrices the same. As a result the observed masses and mixings can be
accommodated in a theory in which there is a very large underlying family and
flavour symmetry group. We have explored the phenomenological implications
of an $SU(3)$ family symmetry together with a $SO(10)$ GUT flavour symmetry
and additional Abelian family symmetries, chosen to restrict the allowed
Yukawa couplings. Allowing for spontaneous (perturbative) breaking of this
group we found a symmetry breaking scheme in which the observed hierarchical
quark masses and mixings are quantitatively described together with the
hierarchy of charged lepton masses and an hierarchical structure for
neutrino masses with near bi-maximal mixing in the lepton sector. In this
the presently unknown mixing angle, $(\theta _{MNS})_{13},$ is determined
mainly by the mixing in the lepton sector. While smaller than the other
mixing angles, it is close to the present limits, and is large enough to
allow for significant CP\ violating effects to be visible in future long
baseline neutrino experiments.

Given the very large underlying symmetry, the fermion masses are heavily
constrained. The perturbative breaking ensures an hierarchical structure for
the masses and in terms of the breaking parameters, the order of magnitude
of the ratios of the quark and lepton masses is determined once the family
symmetry properties of the fields are determined. For quarks, having constrained the family charges to fix the down quark mass ratios, one obtains one order of magnitude prediction for the ratio of  up quark masses which is in good agreement with experiment. Further
the family symmetry ensures an approximate texture zero in the $(1,1)$
matrix element which gives rise to the successful GST relation between the
CKM mixing and the mass ratios of the first two generations of up and down
quarks.

The extension of the family symmetry to leptons can be done together with
via an enlarged GUT flavour symmetry, $SO(10).$ This results in predictions
for the charged lepton masses in terms of the symmetry breaking pattern of $%
SO(10).$ For one particular choice, the
predictions are in good agreement with measurement, reproducing the Georgi
Jarlskog structure for the light lepton masses. In addition the $SO(10)$
symmetry relates the Yukawa couplings of the up and down quark sectors and
replaces the order of magnitude prediction for the ratio of up quarks by an
absolute prediction. For the preferred choice of symmetry breaking this
prediction is in excellent agreement with experiment.

The neutrino mass matrices are also strongly constrained by the symmetry. If
the expansion parameter for the Majorana masses is smaller than that for the
up quarks, the Majorana mass matrix is approximately diagonal in the family
symmetry basis. In this case the lepton mixing angles are determined by the
Dirac mass matrices for the charged leptons and neutrinos and these in turn
are related by the GUT symmetry to the Dirac mass matrices of the quarks. As
a result the lepton mixing angles\footnote{The neutrino masses, while constrained in some case, are largely determined by the unknown Majorana mass eigenvalues.} are determined by the properties of the
quarks. The mixing angle relevant to atmospheric neutrino mixing is well
determined by the leading order operators in the symmetry breaking
expansion. Through vacuum alignment in the $SU(3)$ family sector this mixing
angle is large and consistent with the measure atmospheric mixing
angle. The solar mixing angle turns out to be quite sensitive to subdominant
terms in the symmetry breaking expansion. As a result one obtains only an
order of magnitude prediction for this angle{\it \ }determining it to
be of $O(1).$ For the favoured $SO(10)$ symmetry breaking pattern the
coefficients of the sub-dominant operators needed to obtain a quantitative
agreement of this mixing angle with data are consistent with the constraints
on the expansion which follow from fitting the down quark mass matrix
structure. Finally the value of $(\theta _{MNS})_{13}$ is determined to be
approximately $\sqrt{m_{e}/m_{\mu }}$, coming mainly from the charged lepton
sector.

To summarise, the data on all quark and lepton masses and mixings is
qualitatively consistent with a significantly enlarged family and flavour
symmetry. Given the disparity between these quantities in the quark and
lepton sectors this is already remarkable. On a more quantitative level, a
specific pattern of family symmetries and symmetry breaking leads to a
quantitative prediction for seven of the parameters of the Standard Model,
namely ($\theta _{CKM})_{12},$ $m_{u}m_{t}/m_{c}^{2},$ $m_{e},$ $m_{\mu },$ $%
m_{\tau }$ and $(\theta _{MNS})_{23}$ together with an order of magnitude
prediction for $(\theta _{MNS})_{13}.$ Such agreement is encouraging and
suggests one should try to construct a complete underlying theory, be it a
SUSY GUT or perhaps a superstring theory.\bigskip

\noindent {\large {\bf Acknowledgements}}\bigskip

We would like to thank A.Ibarra, S.King, I.Masina and O.Vives for useful
conversations. L. Velasco-Sevilla would like to thank CONACyT-Mexico and
Universities UK through an ORS Award for financial support.

\bibliographystyle{h-physrev4}
\bibliography{nt}

\appendix

\section{Appendix}


\begin{table}[ht*] \centering%
%
\begin{tabular}{|c|c|c|c|c|}
\hline
& LMA & LOW & VAC &  Ref.\\ \hline
$\tan ^{2}\theta _{12}$ & $(4.0\pm 1)\times 10^{-1}$ & $(5.8\pm 4)\times 10^{-1}$ & $(1.23\pm 0.1)$& \\ 
$\frac{|M_{2}|}{|M_{3}|}^{2}$ & $3\frac{m_{1}^{2}}{m_{2}^{2}}$ & 
$1.77\frac{m_{1}^{2}}{m_{2}^{2}}$ & $8.2\times 10^{-1}\frac{m_{1}^{2}}{m_{2}^{2}}$  &  \\
$(\tan ^{2}\theta _{12})_{{\rm exp}}$ & $4.0\times 10^{-1}$ & $5.8\times
10^{-1}$ & $(0.68,1.8)$ & \cite{pdholsmsno02, bagopesno02,alianisno02} \\ 
$\frac{|\Delta m_{12}^{2}|}{|\Delta m_{23}^{2}|}_{{\rm exp}}$ &
$(0.3,33.6)\times 10^{-2}$ & $(0.48,10)\times 10^{-5}$ & $\approx 1.16\times
10^{-7}$ & \cite{Gonzalez-Garcia:2000sq} \\ 
$\frac{|\Delta m_{12}^{2}|}{|\Delta m_{23}^{2}|}_{{\rm exp}}$ &
$(0.5,30.8)\times 10^{-2}$ & $(0.73,10)\times 10^{-5}$ & $\approx 1.44\times
10^{-7}$ & \cite{Maltoni:2002at} \\ \hline
\end{tabular}
\caption{ Comparision between the predictions for $k=2$ and the
experiments for $\tan^{2}\protect\theta_{12}$ and $|\frac{M_{2}}{M_{3}}|^2$
for the LMA, LOW and VAC solutions for solar neutrinos. The confidence levels
presented are at $3\protect\sigma$.\label{appendixt1}}%
\end{table}%

%

\begin{table}[ht*] \centering%
%
\begin{tabular}{|c|c|c|c|c|c|c|}
\hline
\multicolumn{7}{|c|}{Atmospheric+chooz information} \\ \hline\hline
$\Delta m_{32}^{2}$ &  & $t_{23}^{2}$ &  & $t_{13}^{2}$ &  & Ref. \\ \hline
BFP & Range & BFP & Range & BFP & at 90 \%CL &  \\ \hline
$3.1\times 10^{-3}$ & $(1.1,7.3)\times 10^{-3}$ & $1.4$ & $(0.33,3.7)$ & $%
0.005$ & $<0.055$ & \cite{Gonzalez-Garcia:2000sq} \\ 
& at 99 \%CL &  & at 99 \%CL &  &  &  \\ 
$2.5\times 10^{-3}$ & $(1.2,4.8)\times 10^{-3}$ & $1$ & $(0.43,2.3)$ &  &  & 
\cite{Maltoni:2002at} \\ 
& at $3\sigma $ &  & at $3\sigma $ &  &  &  \\ \hline
\end{tabular}
\caption{The constraints on neutrino mixing parameters
coming from Atmospheric neutrino data and from CHOOZ. The second analysis cited
has been performed with updated atmospheric data. For
different values of $t_{13}^2$ there are different two dimensional C.L.
regions for the variables $t_{23}$ and $\Delta m_{23}^2$; here we present
the appropriate for the value of $t_{13}^2$ obtained with the mass matrix
structure discussed here.\label{appendixt2}}%
\end{table}%
%

\begin{table}[ht*] \centering%
%
\begin{tabular}{|c|c|c|c|c|c|}
\hline
\multicolumn{6}{|c|}{LMA, solar information} \\ \hline\hline
$\Delta m_{12}^{2}$ (eV)$^{2}$ &  & $t_{12}^{2}$ &  & g.o.f. & Ref. \\ \hline
BFP & at $3 \sigma$ CL & BFP & at $3 \sigma$ CL &  &  \\ \hline
$5.5\times 10^{-5}$ & $(0.23,3.7)\times 10^{-4}$ & $4.2\times 10^{-1}$ & $%
(2.4,8.9)\times 10^{-1}$ & 49\% & \cite{bagopesno02} \\ 
$5.6\times 10^{-5}$ & $(0.22,2.2)\times 10^{-4}$ & $3.9\times 10^{-1}$ & $%
(2.0,6.4)\times 10^{-1}$ &  & \cite{bamawhisno02} \\ 
$6.2\times 10^{-5}$ & $(0.23,3.4)\times 10^{-5}$ & $4.0\times 10^{-1}$ & $%
(2.3,7.9)\times 10^{-1}$ & 84\% & \cite{pdholsmsno02} \\ \hline
& at $ 1\sigma$ CL &  & at $1\sigma$ CL&  &  \\ 
$4.5\times 10^{-5}$ & $(3.1,7.2)\times 10^{-1}$ & $4.0\times 10^{-1}$ & $%
(3.2,4.8)\times 10^{-1}$ &  & \cite{alianisno02} \\ \hline
\end{tabular}
\caption{ Experimental constraints for LMA solution. BFP
refers to the best fit point given in the cited references, the $3\protect\sigma$=$99.73\%$ CL or the $1\protect\sigma$ C.L. is shown. \label{appendixt3}}%
\end{table}%
%

\begin{table}[tbp]
\label{lows}
\par
\begin{center}
\begin{tabular}{|c|c|c|c|c|c|}
\hline
\multicolumn{6}{|c|}{LOW, solar information} \\ \hline\hline
$\Delta m_{12}^{2}$ & $\Delta m_{12}^{2}$ & $t_{12}^{2}$ & $t_{12}^{2}$ & 
g.o.f. & Ref. \\ \hline
BFP & at $3 \sigma $ CL & BFP & at $3 \sigma $  CL &  &  \\ \hline
$1.0\times 10^{-7}$ & $(0.5,1.1)\times 10^{-7}$ & $7.1\times 10^{-1}$ & $%
(5.5,10)\times 10^{-1}$ & 45\% & \cite{Bahcall:2001zu} \\ 
$1.1\times 10^{-7}$ & $(0.6,1.2)\times 10^{-7}$ & $6.9\times 10^{-1}$ & $%
(4,9)\times 10^{-1}$ & 69\% & \cite{Krastev:2001tv} \\ 
& $(0.9,1)\times 10^{-7}$ &  & $(6,7.5)\times 10^{-1}$ &  & \cite
{Fogli:2001vr} \\ \hline
\end{tabular}
\end{center}
\caption{{\protect\small {Experimental constraints for the LOW solution. BFP
refers to the best fit point given in the cited references, the $3 \sigma$ CL
has been estimated from those references. The analyses cited have been
performed using the latest SNO  information.}}}
\label{appendixt4}
\end{table}

\begin{table}[tbp]
\label{vacs}
\par
\begin{center}
\begin{tabular}{|c|c|c|c|c|c|}
\hline
\multicolumn{6}{|c|}{VAC} \\ \hline\hline
$\Delta m_{12}^{2}$ & $\Delta m_{12}^{2}$ & $t_{12}^{2}$ & $t_{12}^{2}$ & 
g.o.f. & Ref. \\ \hline
BFP &at $3 \sigma$ CL  & BFP & at $3 \sigma$ CL &  &  \\ \hline
$4.6\times 10^{-10}$ & $(3.5,5.7)\times 10^{-10}$ & $2.4\times 10^{0}$ & $%
(0.3,3.5)$ & 42\% & \cite{Bahcall:2001zu} \\ \hline
\end{tabular}
\end{center}
\caption{{\protect\small {Experimental constraints for VAC solution. BFP
refers to the best fit point given in the cited references, the $3 \sigma$ CL
has been estimated from those references. The analysis cited has been
performed using the latest SNO information and with an enhanced CC cross section for deuterium, as quoted in \protect\cite{Bahcall:2001zu}.}}}
\label{appendixt5}
\end{table}

\end{document}